\documentclass[pra,showpacs,superscriptaddress,floatfix,reprint]{revtex4-1}

\usepackage{amsmath}
\usepackage{graphicx}  

\usepackage[update,prepend,prefersuffix=false,suffix=]{epstopdf}

\begin{document}

\title{Localization from quantum interference in one-dimensional disordered potentials}

\author{M. Moratti}
\affiliation{LENS, Universit\`a di Firenze, Via N. Carrara 1, 50019 Sesto Fiorentino, Italy}
\author{M. Modugno}
\affiliation{Department of Theoretical Physics and History of Science, UPV-EHU, 48080 Bilbao, Spain}
\affiliation{IKERBASQUE, Basque Foundation for Science, 48011 Bilbao, Spain}

\date{\today}

\begin{abstract}
We show that the tails of the asymptotic density distribution of a quantum wave packet that localizes in the the presence of random or quasiperiodic disorder can be described by the diagonal term of the  projection over the eigenstates of the disordered potential. This is equivalent of assuming a phase randomization of the off-diagonal/interference terms. We demonstrate these results through numerical calculations of the dynamics of ultracold atoms in the one-dimensional speckle and quasiperiodic potentials used in the recent experiments that lead to the observation of Anderson localization for matter waves [Billy \textit{et al.}, Nature {\bf 453}, 891 (2008); Roati \textit{et al.}, Nature {\bf 453}, 895 (2008)]. For the quasiperiodic case, we also discuss the implications of using continuos or discrete models.
\end{abstract}

\maketitle

Since the seminal work by P. W. Anderson \cite{anderson}, the behavior of quantum particles in disordered potentials has been a central topic in physics \cite{lifshits,kramer}. In particular, the diffusion of a quantum wave packet in a one-dimensional disordered systems has been longly investigated from the theoretical point of view \cite{zhong,flach,sp,succi,larcher,shep,salasnich,albert,piraud}, and also observed experimentally in two recent experiments with matter waves \cite{billy,roati}. In these experiments,  the initial wave packet is prepared as the ground state of a trapped Bose-Einstein condensate. Then, the trapping is switched off and the condensate is let expand in a one-dimensional wave guide in the presence of disorder, in a regime where interatomic interactions can be neglected. Under certain conditions, disorder induced exponential localization (Anderson localization) is observed. Similar results have been reported for the propagation of light  in quasiperiodic photonic lattices \cite{lahini}.

Usually, the localization is interpreted in terms of localization of single plane wave components, as a result of multiple scattering from the random potential \cite{sp,billy}. Recently, a more elaborate approach has been discussed in \cite{piraud}.
In this article we propose an alternative and complementary viewpoint, in terms of the eigenstates of the disordered potential $V(x)$. We show that the tails of the disorder averaged asymptotic density distribution are determined by the diagonal term of the projection of the initial wave packet on the eigenstates of the single particle Hamiltonian $H={p^2}/{2m} +V(x)$, and that this is equivalent to an effective randomization of the eigenstate phases. In fact, though the interference term does not vanish (actually, its amplitude is always of the order of the diagonal term), it contributes just with local fluctuations.

We demonstrate these results through numerical work for the one-dimensional speckle and quasiperiodic potentials employed in recent experiments with ultracold atoms \cite{billy,roati}. This analysis highlights the role of quantum interference in the localization process,  demonstrating that it plays a key role for both correlated disorder and quasiperiodic potentials, even in the deep localizing regime. For the quasiperiodic case, we also discuss the different features of continuos and discrete models.

\textit{Model.}
Let us consider a one-dimensional non-interacting system, namely a particle of mass $m$ in a  potential $V(x)$, described by the hamiltonian $H=-({\hbar^2}/{2m})\nabla^2_{x} +V(x)$. For simplicity we consider a finite size system of length L, whose discrete spectrum is characterized by eigenvalues $E_{n}$ and eigenvectors $\phi_{n}(x)=\langle x|n\rangle$. 
Then, the evolution of an initial wave packet $\langle x|\psi(0)\rangle$ can be expressed in terms of the eigenstates $|n\rangle$ as
\begin{equation}
\langle x|\psi(t)\rangle = \sum_{n} c_{n}\langle x|n\rangle e^{-i E_{n}t/\hbar},
\label{eq:simple-projection}
\end{equation}
with $c_{n}=\langle n|\psi(0)\rangle$. 
By writing the projection coefficients in the modulus-phase representation, $c_n\equiv|c_n|\exp(i\theta_{n})$, we can rewrite Eq. (\ref{eq:simple-projection}) as
$\langle x|\psi(t)\rangle = \sum_{n} |c_{n}|\phi_{n}(x) e^{i \varphi_{n}(t)}$, with $\varphi_{n}(t)=\theta_{n}-E_{n}t/\hbar$.
These equations are equivalent to solving the Schr\"odinger equation
$i\hbar\partial_t\psi(x)=H\psi(x)$,
but have the advantage that
the density profile at any time can be computed directly on the knowledge of just the initial conditions.
 The density distribution $n(x,t) \equiv |\langle x|\psi(t)\rangle|^{2}$ at any time $t$ is given by
\begin{align}
n(x,t) =&\sum_{n} |c_n|^2|\phi_{n}(x)|^2 \nonumber\\
&+ \sum_{n\neq n'} |c_n| |c_{n'}|\phi_{n}^{*}(x)\phi_{n'}(x)e^{i \Delta\varphi_{nn'}(t)}
\nonumber\\
&\equiv n_{diag}(x) + n_{off}(x,t)
\label{eq:free-evolution}
\end{align}
where we have defined the
phase difference $\Delta\varphi_{nn'}(t)\equiv \varphi_{n}(t)-\varphi_{n'}(t)$, and explicitly separated the diagonal and the off-diagonal interference terms.
The evolution of the latter is determined by the phase difference $\Delta\varphi_{nn'}(t)$, whereas the diagonal term does not depend on time.

Conceptually, in the case of disorder, this approach is complementary to the usual description of localization in terms of the localization of single momentum components (plane waves), as a result of multiple scattering from the random potential \cite{sp}. Its advantages are that each component evolves independently from the others (instead, plane waves are not independent - even in the non-interacting case - because they are coupled by the potential, see also \cite{piraud}), and that the time evolution of each component can be simply obtained from the knowledge of the energy spectrum $E_{n}$. From the practical point of view, this allows to compute the wave-function at arbitrary long times with the effort of a single time-step. 

In the limit $L\to\infty$ the spectrum may become continuous, or remain discrete, depending on the potential $V(x)$. In the first case, it is known that the initial wave packet will spread indefinitely, whereas this cannot happen if the spectrum is discrete \cite{landau}. The discreteness of the spectrum can derive both from the presence of a confining potential, and in this case the boundedness of the dynamics is a trivial consequence, or more interestingly due to the presence of disorder \cite{lifshits}. The implications of this latter case will be discussed in the following.

\textit{Asymptotic regime.}
Let us consider the effect of the interference term in Eq. (\ref{eq:free-evolution}), in the asymptotic (long times) regime. 
Initially, the phase relations between different components are exactly those that reconstruct the assigned initial profile. In case of a real wave packet, $\theta_{n}=0$ or $\pi$. 
Then, they evolve in a deterministic way, according to $\varphi_{n}(t)=\theta_{n}-E_{n}t/\hbar$, determining the wave packet profile as a result of the interference of many components.  
For a disordered system in the localized regime - characterized by a \textit{point} (discrete) spectrum  - we expect the asymptotic  distribution, $n_{\infty}(x)=\lim_{t\to \infty}n(x,t)$, to be stationary (except for local fluctuations) and determined by the localization properties of the eigenstates \cite{lifshits}. 

We argue here that, for a disordered system in the localized regime, the behavior of the tails of the asymptotic density distribution is determined just by diagonal terms
\begin{equation}
n_\infty^{tails}(x) \approx \sum_{n} |c_n|^2|\phi_{n}(x)|^2
\label{eq:asymptotic_density}
\end{equation}
and can be therefore computed by the knowledge of only the initial coefficients $c_{n}$ and the localization properties of the eigenstates $\phi_{n}(x)$. In fact, it turns out that the interference (off-diagonal) term asymptotically contributes just with fluctuations around the profile of Eq. (\ref{eq:asymptotic_density}). This can be explained in terms of an effective randomization of the phases $\varphi_n$, as will be discussed later on. 
The approximate equivalence in Eq. (\ref{eq:asymptotic_density}) refers to local fluctuations related to the specific disorder realization, the relation becoming more and more accurate when averaged over many disorder realizations.

To characterize the relaxation of the the full density on the diagonal term during the evolution of the system, we introduce the following quantity
\begin{equation}
R(t)\equiv\frac{1}{L}\int_{L}\!\!dx\frac{|\ln(n_{diag}(x))|}{|\ln(n(x,t))|}
\label{eq:rx}
\end{equation}
that measures the average relative weight of the diagonal term (in logarithmic scale). 
This definition is especially useful when the density is exponentially localized.
Asymptotically, we expect $R(t)$ to approach 1 if $n_{diag}(x)$ is a good approximation of $n(x,t)$. 

As far as concerns the shape of the asymptotic density distribution, when the eigenstates are localized around $x_{n}=\langle \phi_n|x| \phi_n\rangle$
and decay exponentially away from the localization center, $|\phi_{n}(x)|^2\approx \exp[-(x-x_{n})|/\ell_{n}]$, the long distance behavior of the asymptotic wave packet is determined by the largest localization length $\ell=max\{\ell_{n}\}$. However, the overall profile may not be characterized by a simple exponential decay, due to the superposition of many energy components with different localization lengths \cite{sp,piraud}.

In the following we illustrate these ideas for the localization of matter waves, as considered in recent experiments \cite{billy,roati}. In both cases the spectrum  is computed by mapping the stationary Schr\"odinger equation $H\phi_{n}=E_{n}\phi_{n}$ on a discretized grid \cite{numerics}. All the results presented in this paper are averaged over 100 realizations of disorder.

\textit{Quasiperiodic potential.}
Let us start by discussing the case of a quasiperiodic potential as considered in
\cite{roati,larcher,albert}. The quasiperiodic potential is created by superimposing two optical lattices, of wavevector $k_{i}$ and intensity $V_{i}$ ($i=1,2$), $V(x)=V_{1}\sin^{2}(k_{1}x) + V_{2}\sin^{2}(k_{2}x + \theta)$, with $V_{1}\gg V_{2}$, corresponding to a primary optical lattice weakly modulated by the second one, $\theta$ being an arbitrary phase. 
In this case, it is convenient to express lengths in units of the lattice spacing  $d=\pi/k_{1}$ and energies in terms of the associated recoil energy  $E_{R1}=\hbar^{2}k_{1}^{2}/2m$.
In the tight binding regime, $V_{1}\gg E_{R1}$, the system Hamiltonian can be mapped onto the discrete Aubry-Andr\'e model \cite{aubry} for the particle amplitude $\varphi_j$ at site $j$ of the primary lattice
$H=\sum_j-J(\varphi_{j+1} \varphi_j^*+c.c.)+\Delta\cos(2\pi\beta j+\theta)| \varphi_j|^2$,
where $\beta=k_2/k_1$ is the ratio between the two lattice wavevectors, and the tunneling constant $J$ and disorder amplitude $\Delta$ are functions of the amplitudes $V_{i}$  \cite{boers,modugno09}.
When the two wavevectors are incommensurate, the model is characterized by a transition from extended to localized states, at $\Delta/J=2$ \cite{aubry}.
Here we chose $\beta=(\sqrt{5}-1)/2$, the inverse of the golden mean \cite{ingold}, and $\Delta/J=7$, in order to discuss a specific example where localization occurs.
 
As initial state we consider a gaussian wave-packet modulated by the primary lattice, prepared as the non-interacting ground state of the primary lattice ($V_{2}=0$) plus an additional harmonic trapping (here we consider $\omega_{ho}=0.03\, E_{R1}/\hbar$), as in the experiments \cite{roati}. This choice guarantees that the initial state $|\psi(0)\rangle$ has mainly components in first Bloch band of the primary lattice. In addition, in order to make a direct comparison with the solution of the discrete AA model \cite{larcher} - that is intrinsically a single band model - we also consider a truncated initial state $|\bar{\psi}(0)\rangle$, obtained by removing the components lying in higher Bloch bands.

\begin{figure}[t]
\centerline{
\includegraphics[width=0.48\columnwidth]{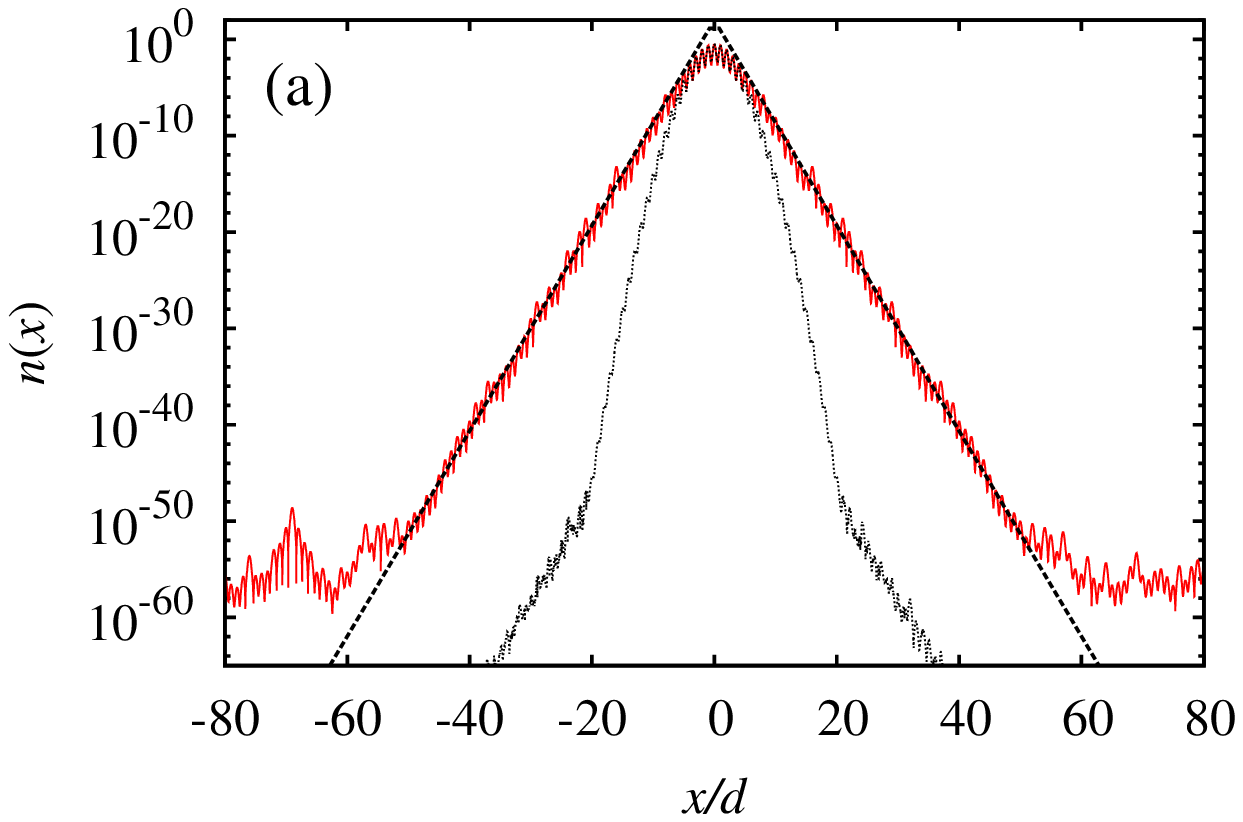}
\includegraphics[width=0.48\columnwidth]{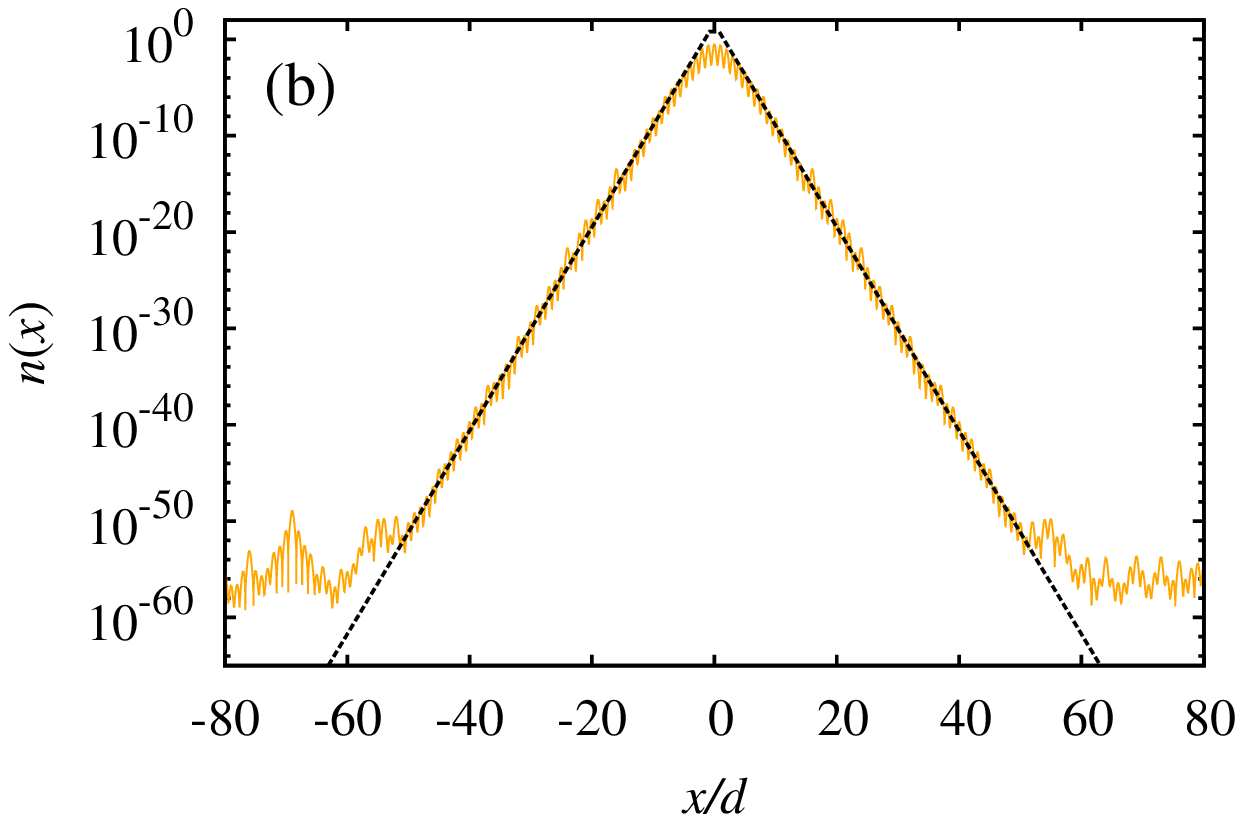}}
\centerline{
\includegraphics[width=0.48\columnwidth]{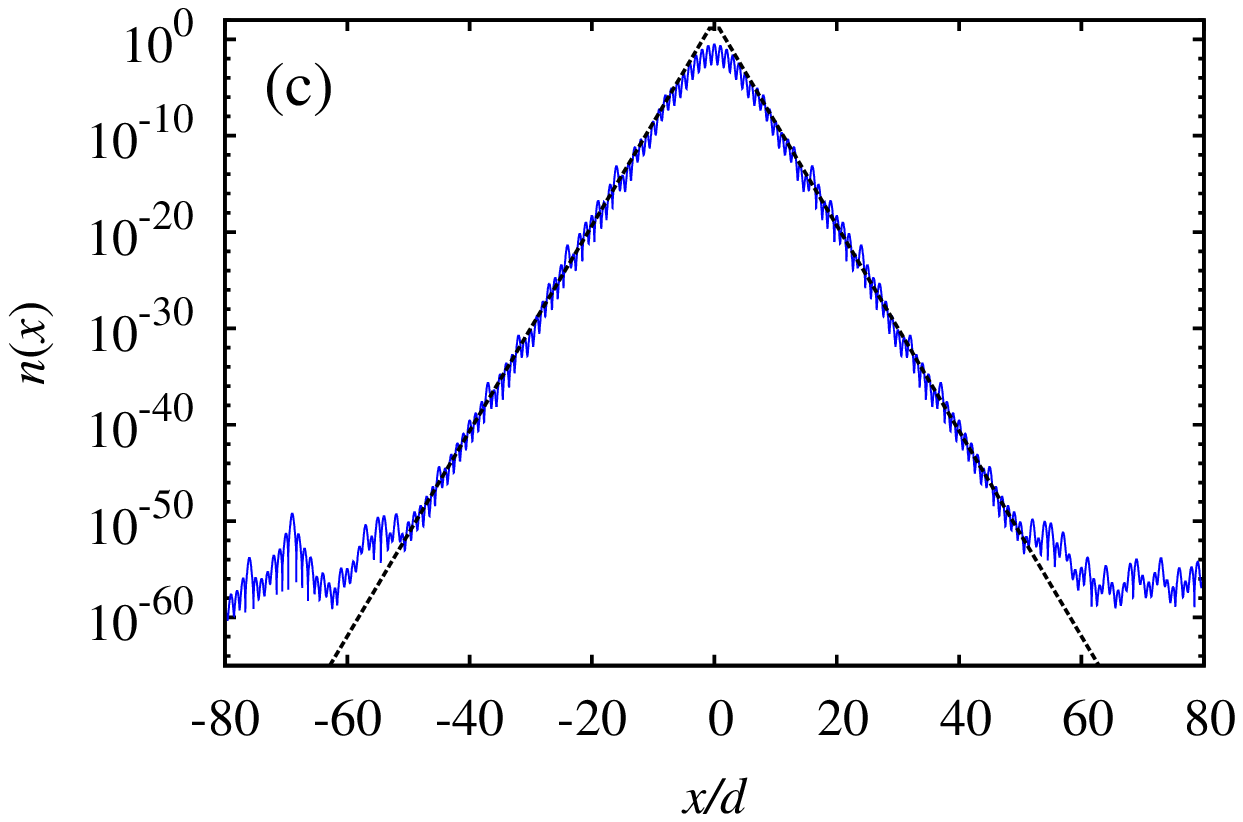}
\includegraphics[width=0.48\columnwidth]{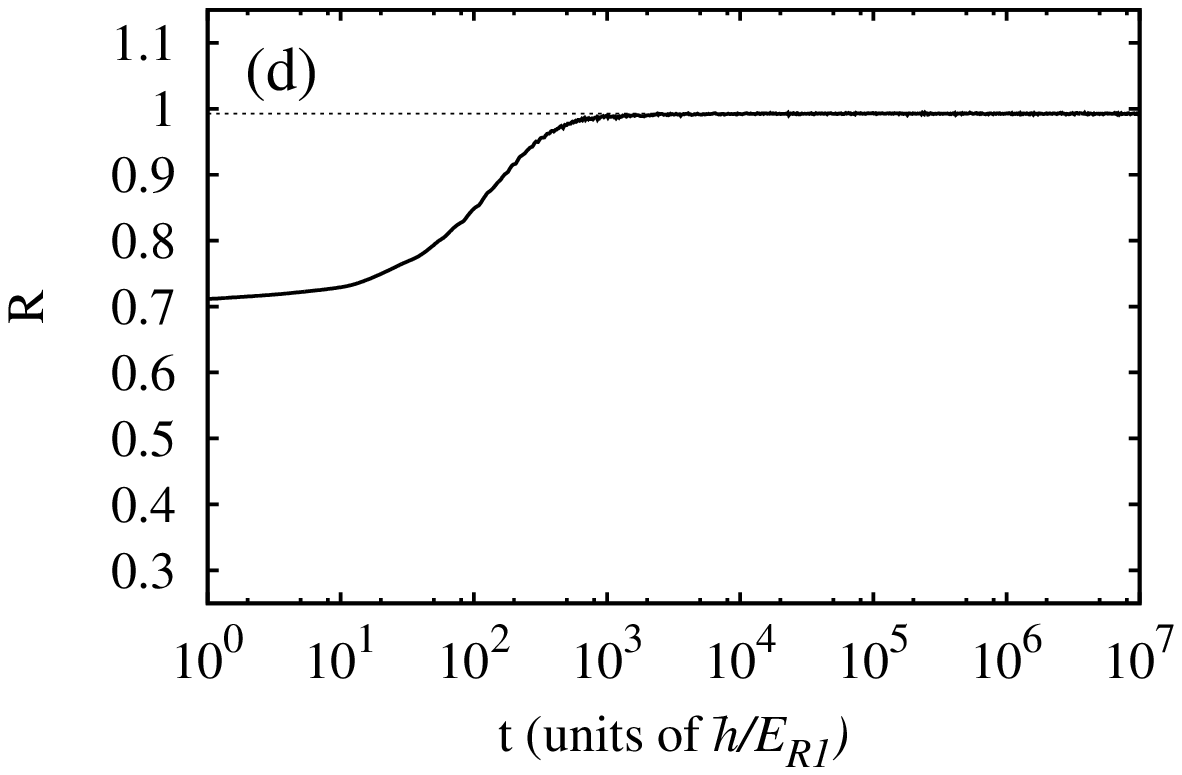}}
\caption{(Color online) (a),(b),(c) Asymptotic density distribution in the quasiperiodic potential, in the localized regime. The full expression (\ref{eq:free-evolution}) (red, solid line in (a)) at $t=10^{7}\, \hbar/E_{R1}$ is compared with the expression in (\ref{eq:asymptotic_density}) (orange, (b)), and 
with the same Eq. (\ref{eq:free-evolution}) but in case of a random choice of the phases $\varphi_{n}$ (blue line, (c)). All the data refer to an initial wavepacket in the \textit{single-band approximation}, also shown in (a) (black dotted line). The straight (dashed) lines in all these three panels are obtained from an exponential fit of the density tails in (a), with a function $\propto \exp[-2|x|/\ell_{loc}]$ (here $\ell_{loc}\approx0.81 d$, see text). 
(d) Evolution of the quantity $R(t)$ in Eq. (\ref{eq:rx}) as a function of time; the dotted horizontal line represents the case of Eq. (\ref{eq:free-evolution}) for a random choice of the phases $\varphi_{n}$ ($R\simeq0.993$).}
\label{fig:quasip-sb}
\end{figure}

Let us start from the case of the truncated initial state $|\bar{\psi}(0)\rangle$. In Fig. \ref{fig:quasip-sb}(a) we show the initial wave packet $n_{0}(x)=|\langle x|\bar{\psi}(0)\rangle|^{2}$ and the average asymptotic density distribution obtained from Eq. (\ref{eq:free-evolution}) at $t=10^{7}\, \hbar/E_{R1}$. The tails of the asymptotic density are characterized by an exponential decay, as found from the solution of the discrete AA model \cite{larcher}. 
In particular, by fitting the density with a function $\propto \exp[-2|x|/\ell_{loc}]$, we find $\ell_{loc}\approx0.81\, d$, that is consistent with the expected value $\ell_{AA}=d/\log(\Delta/2J)|_{\Delta/J=7}\approx0.8\, d$ for the AA model \cite{aubry}. 
Then, in Fig. \ref{fig:quasip-sb}(b) we plot the diagonal term $n_{diag}(x)$, and in Fig. \ref{fig:quasip-sb}(c) again the full expression of  Eq. (\ref{eq:free-evolution}), but in case of a random choice of the phases $\varphi_{n}$. In both cases we keep the same fitting lines of Fig. \ref{fig:quasip-sb}(a) for comparison. These figures show that both approximations in (b) and (c) nicely reproduce the actual density distribution, confirming that Eq. (\ref{eq:asymptotic_density}), or equivalently the {\it phase randomization ansatz}, correctly describes the asymptotic behavior of the tails in the localized regime. Indeed, the three densities in (a), (b) and (c) would be hardly distinguishable if drawn together on the same plot. From Figs. \ref{fig:quasip-sb}(a) and (b) it is easy to evince that the off-diagonal interference terms are essential in determining the evolution of the initial wave packet towards the asymptotic density distribution, but once the system is in the asymptotic regime they can be disregarded, as long as the behaviour of the tails is concerned \cite{offdiag}.  The approach to the asymptotic regime can be tracked by means of the quantity $R(t)$ in Eq. (\ref{eq:rx}), whose evolution is shown in Fig. \ref{fig:quasip-sb}(d). It is characterized by a smooth crossover towards $R\simeq1$ around a ``critical'' time $t_{a}=10^{3}\, \hbar/E_{R1}$, signaling that for $t> t_{a}$ the system has almost relaxed to the asymptotic distribution \cite{DJ}. Indeed, the behavior shown in Figs. \ref{fig:quasip-sb}(a)-(c) for $t=10^{7}\, \hbar/E_{R1}$ is found at any time in the asymptotic regime. Remarkably, we find the same behavior even for a single realization.

\begin{figure}[t!]
\centerline{
\includegraphics[width=0.48\columnwidth]{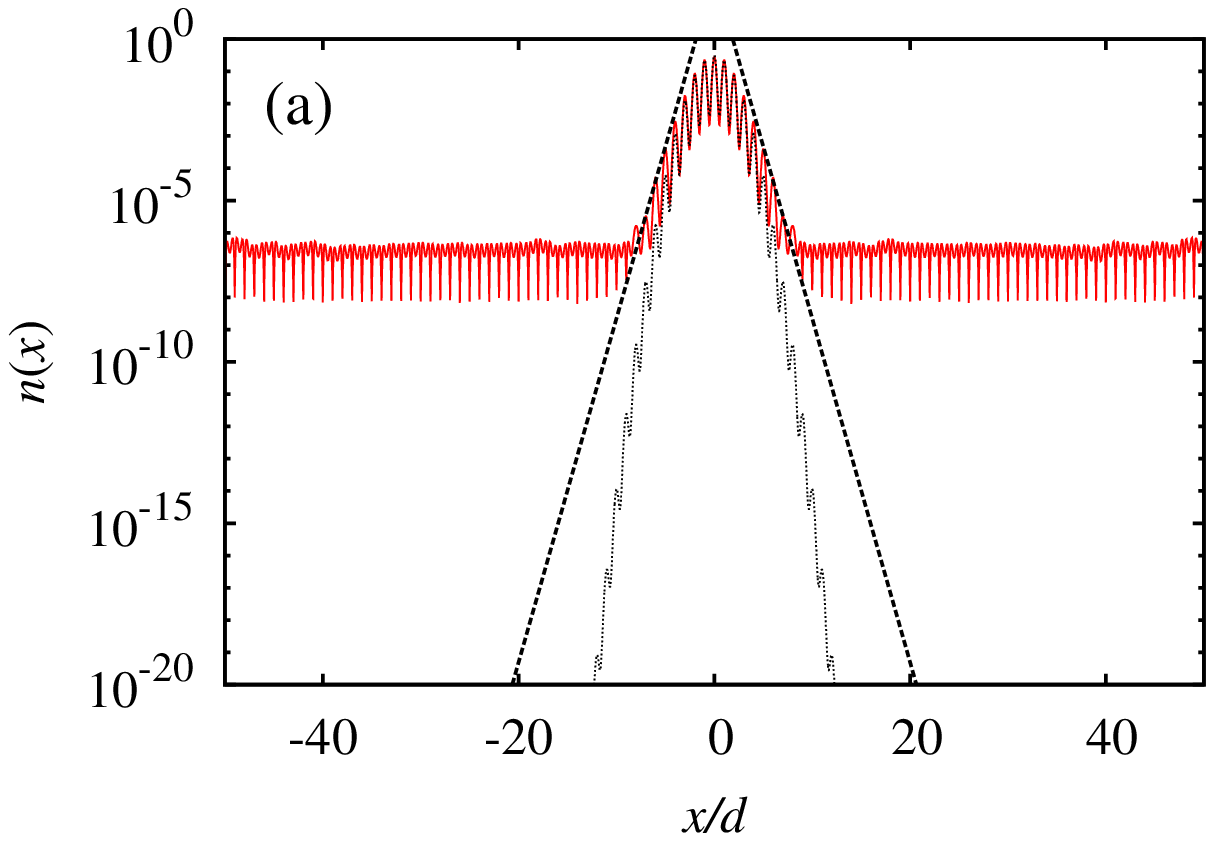}
\includegraphics[width=0.48\columnwidth]{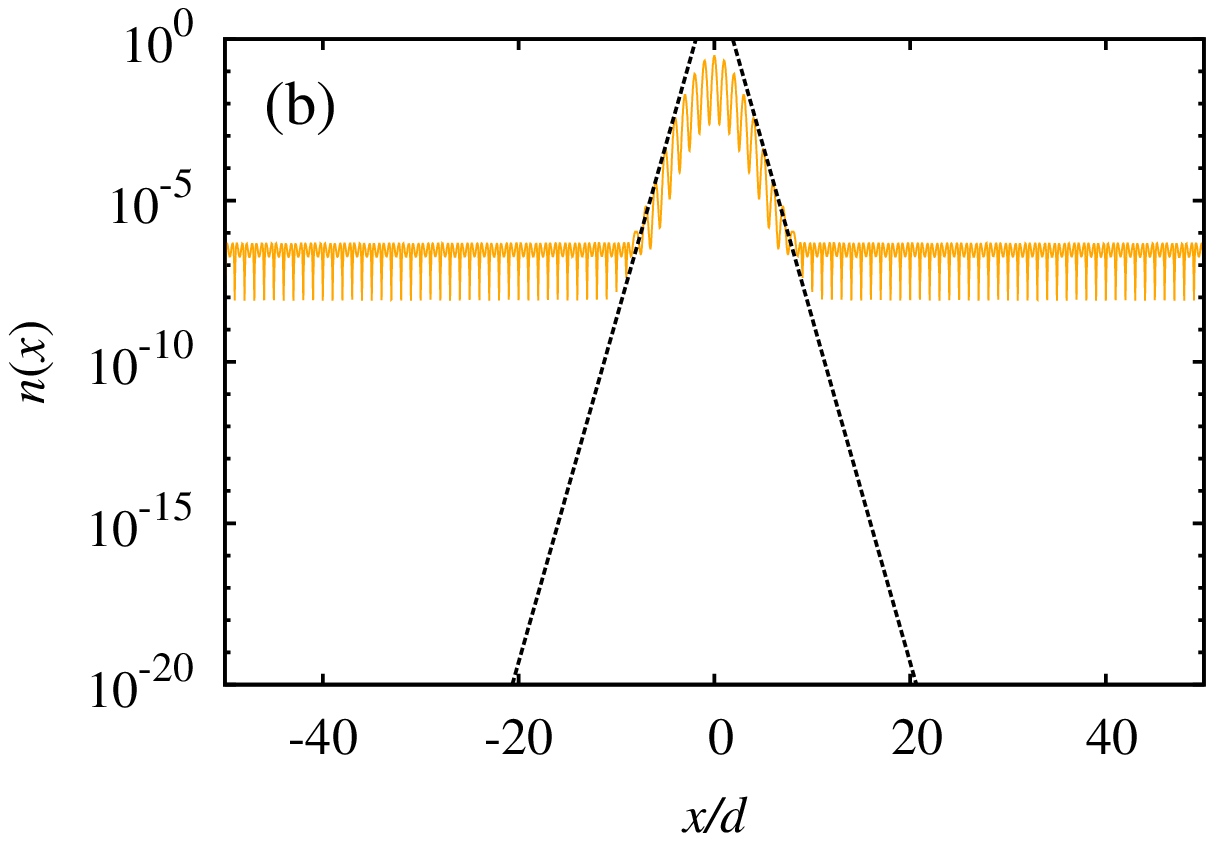}}
\centerline{
\includegraphics[width=0.48\columnwidth]{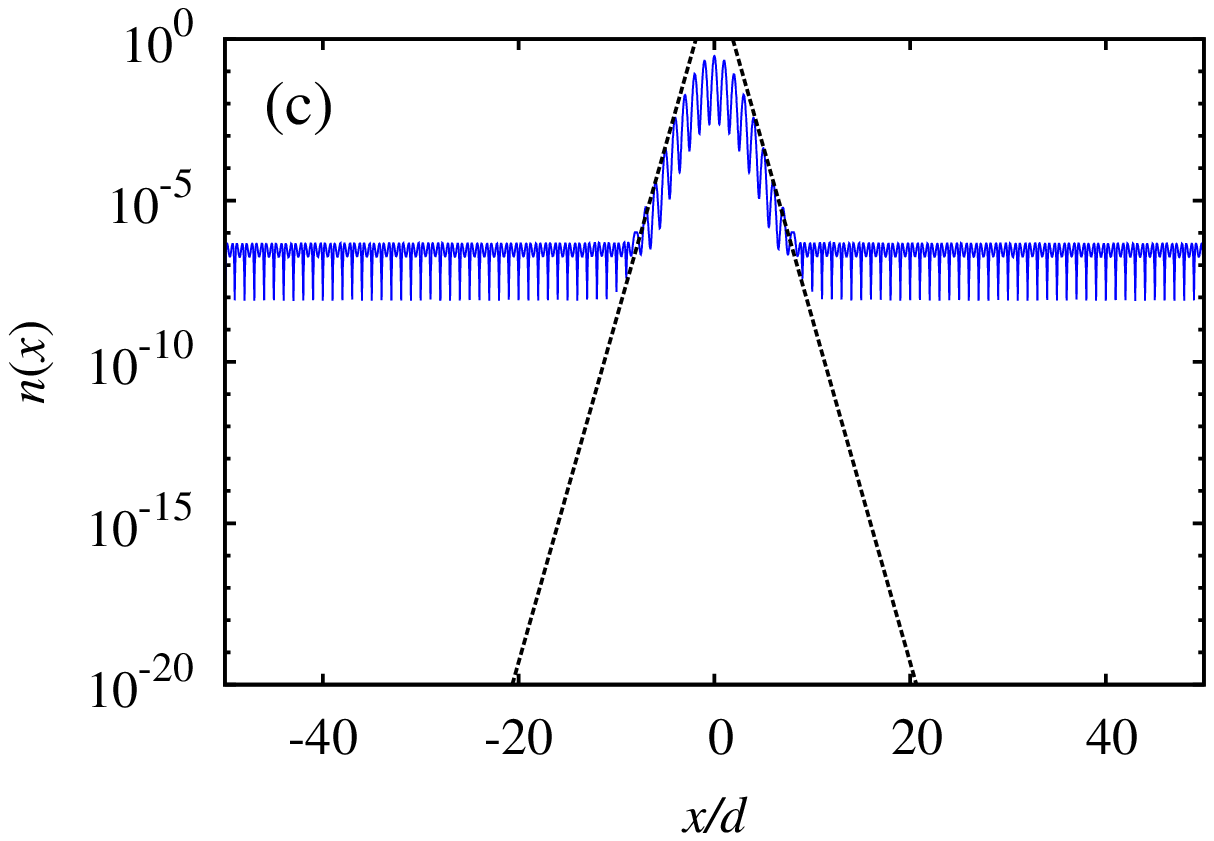}
\includegraphics[width=0.48\columnwidth]{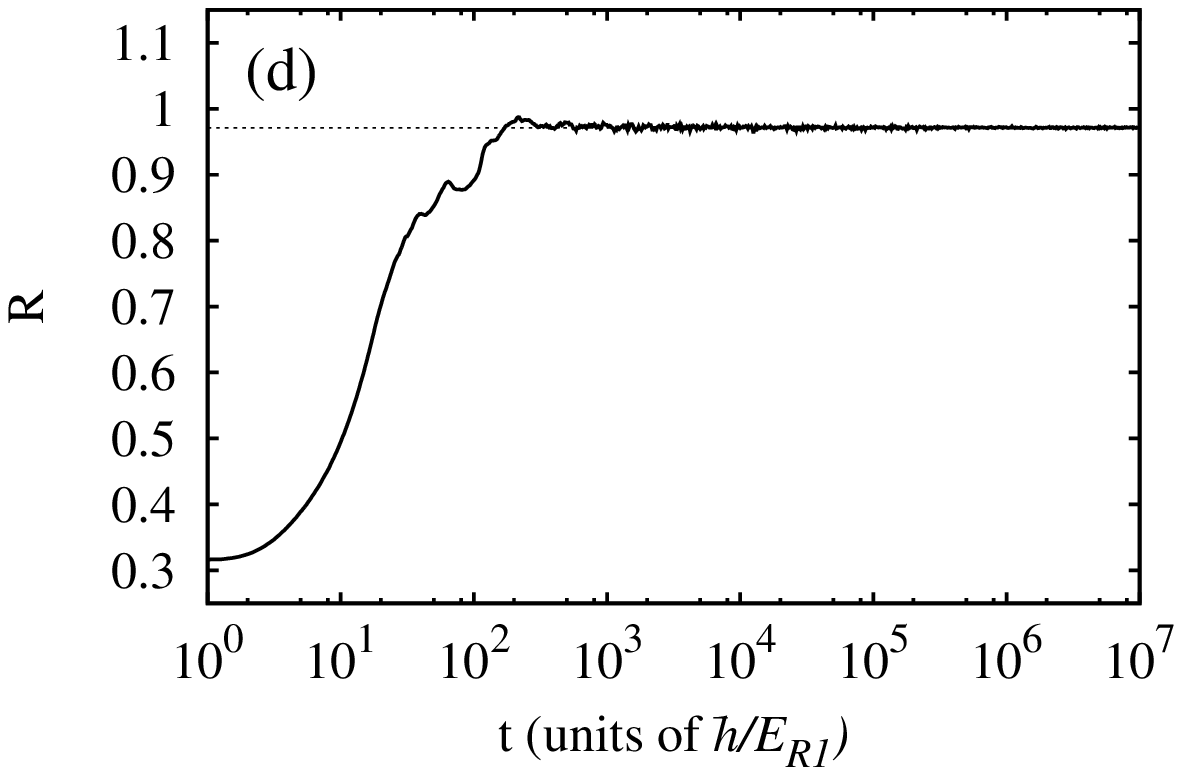}}
\caption{(Color online) Same as Fig. \ref{fig:quasip-sb}, but for the full initial wave packet $\langle x|\psi(0)\rangle$, with components also in excited Bloch bands (black dotted line, in (a)).
(a) Asymptotic density distribution from the full expression (\ref{eq:free-evolution}) (red, solid line); (b) expression in (\ref{eq:asymptotic_density}) (orange); (c) Eq. (\ref{eq:free-evolution}) but in case of a random choice of the phases $\varphi_{n}$ (blue). The exponential fit $\approx \exp[-2|x|/\ell_{AA}]$ (dashed line) is the one obtained from Fig. \ref{fig:quasip-sb}(a). (d) Evolution of the quantity $R(t)$ in Eq. (\ref{eq:rx}) as a function of time, compared with the case of Eq. (\ref{eq:free-evolution}) for a random choice of the phases $\varphi_{n}$ (dotted line, $R\simeq0.971$).}
\label{fig:10bands}
\end{figure}

In order to discuss the effect of higher bands, that are naturally included in the continuous description of real experiments, we now consider the full initial wave packet $\langle x|\psi(0)\rangle$ \cite{tenbands}. The corresponding asymptotic distribution are shown in Figs. \ref{fig:10bands}(a)-(c) (for comparison we also show the same fitting lines of Fig. \ref{fig:quasip-sb}(a); note the different scales between the two figures). With respect to the previous case of a truncated initial state $|\bar{\psi}(0)\rangle$, there is now a continuous background due to the fact that the eigenstates in higher Bloch bands are extended. Nevertheless, even in this case the full solution is correctly described by the asymptotic expression (\ref{eq:asymptotic_density}).
We note that the presence of this background represents an intrinsic cut-off on the extension of   
the exponential tails, that is inherent in the continuous description of the system. Fig. \ref{fig:10bands}(d) shows that in this case the behaviour of $R(t)$, though similar to the previous one, presents some quantitative differences: initially $R(t)$ starts from a lower value, and then saturates slightly below the single band case. Both these effects can be attributed to the presence of the extended states. In particular, the relatively low initial value is due to the very different shape of the initial and asymptotic density distributions.

\textit{Speckle potential.} 
Let us now turn to the case of an optical speckle potential, that is a correlated disorder potential characterized by a negative-exponential distribution of the intensity 
$P(V_{0})=\textrm{e}^{-V_{0}/\langle V_{0}\rangle}/\langle V_{0}\rangle$, 
and a spatial (auto)correlation function $\Gamma(x)= 1 + \textrm{sinc}(0.88 x/\xi)^2$, $\xi $ being the correlation length \cite{goodman,modugno06}. It can be written as $V(x)=V_0 v(x)$
with the distribution of intensities of $v(x)$ being normalized to
unit standard deviation. The speckles are characterized by the presence of an effective mobility edge \cite{sp}, below which the eigenstates localize exponentially. 
Lengths and energies can be conveniently expressed in units of the natural scales provided by the correlation length $\xi$ and its associated energy  $E_{\xi}=\hbar^{2}/2m\xi^{2}$ \cite{falco}.
The speckles can be generated numerically by following the method described in \cite{huntley,modugno06}.

\begin{figure}
\centerline{
\includegraphics[width=0.48\columnwidth]{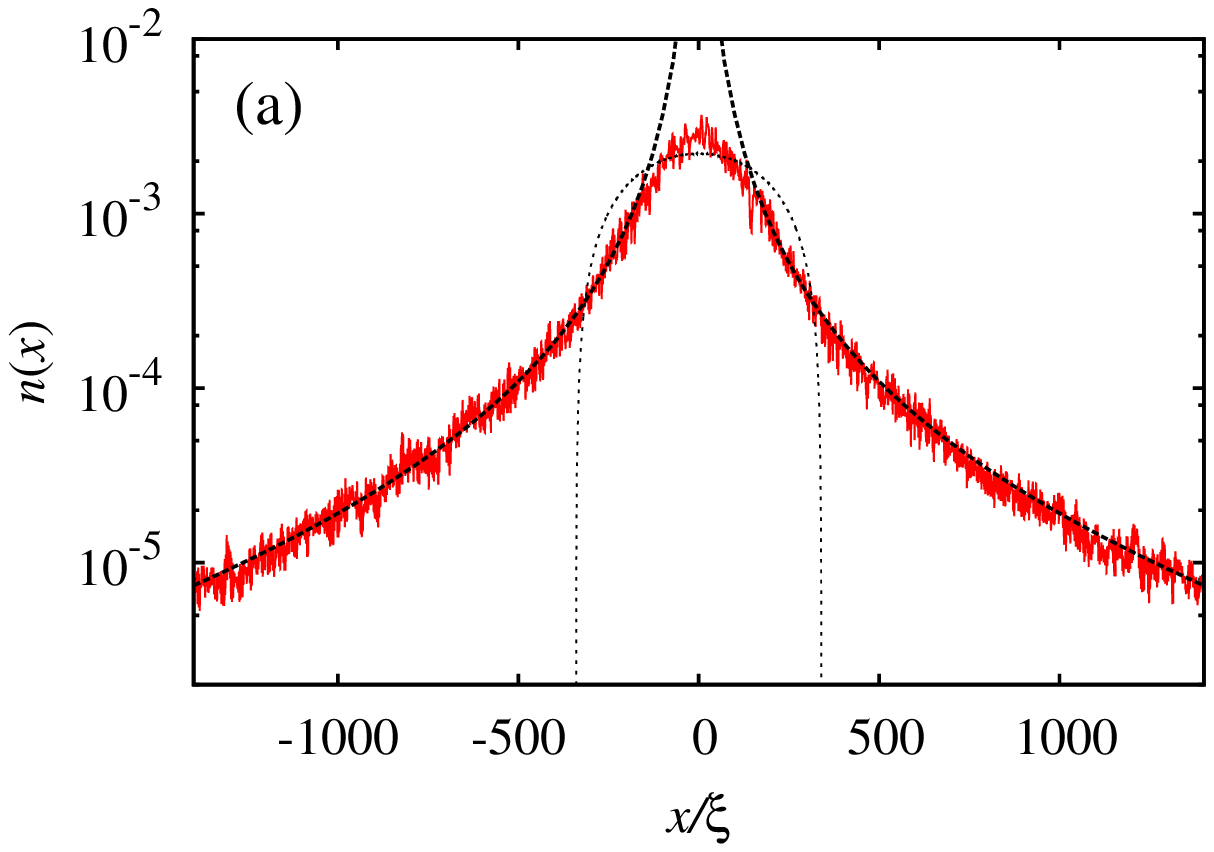}
\includegraphics[width=0.48\columnwidth]{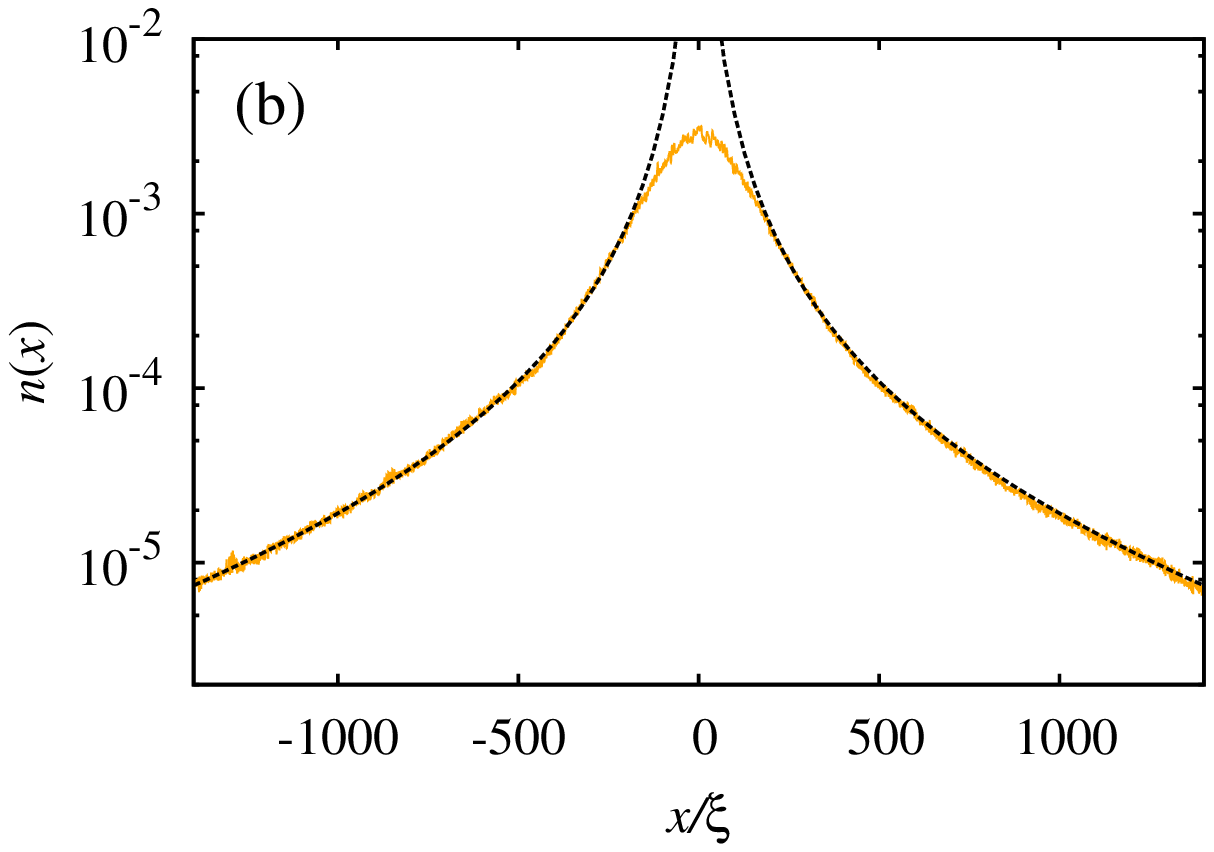}}
\centerline{
\includegraphics[width=0.48\columnwidth]{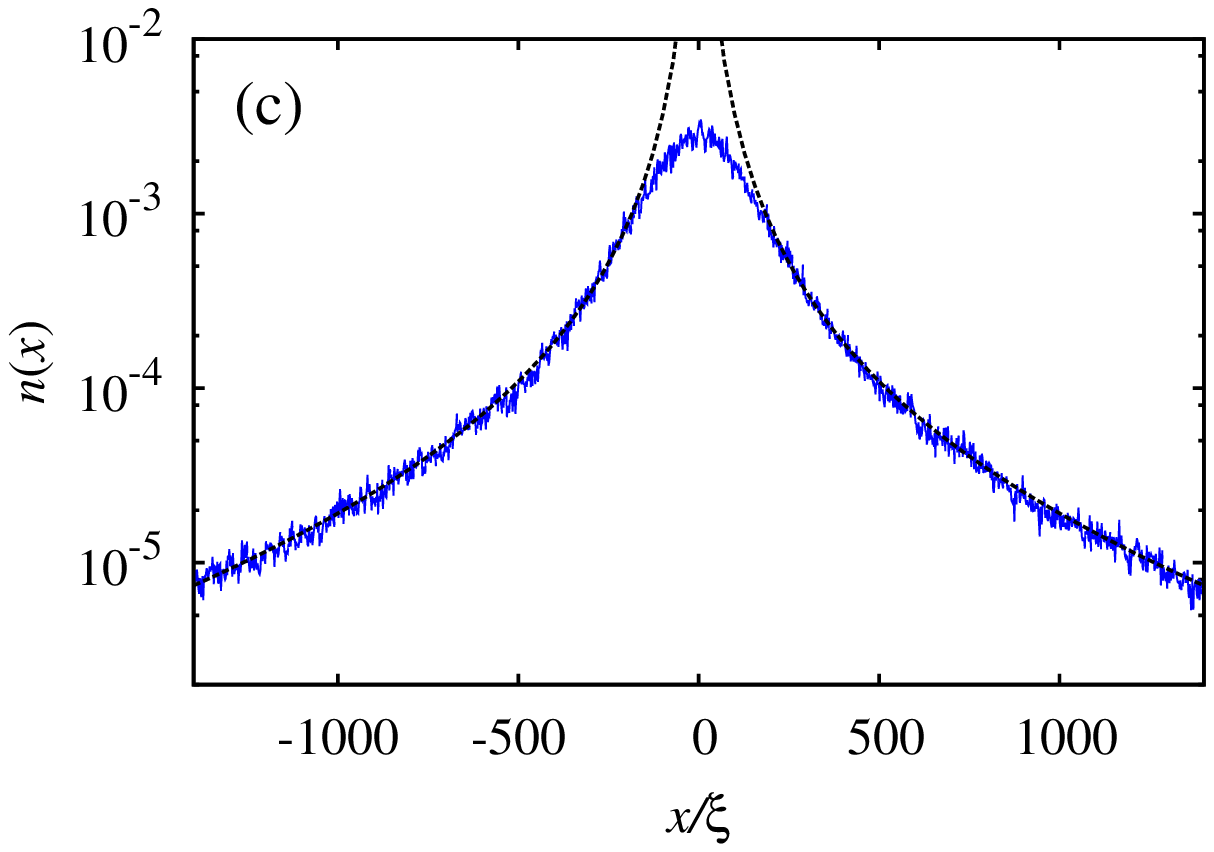}
\includegraphics[width=0.48\columnwidth]{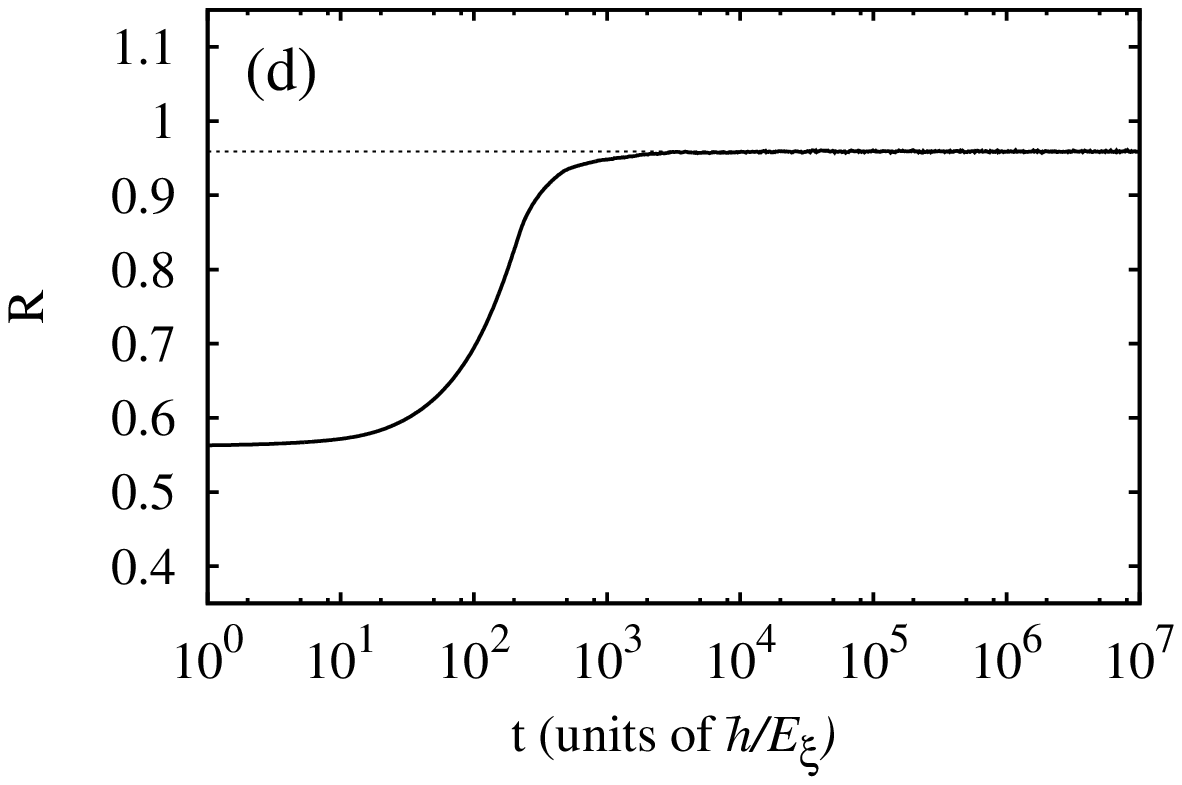}}
\caption{(Color online) Localization in the speckle potential. (a) Initial wave packet (black dotted line) and symptotic density distribution obtained from the full expression in Eq. (\ref{eq:free-evolution}) (red, solid line), with the tails fitted as $\propto|x|^{\alpha}e^{-2\gamma|x|}$ (black, dashed line); the same lines are plotted also in panels (b) and (c) for comparison. (b) Contribution of the diagonal terms, see Eq.  (\ref{eq:asymptotic_density}) (orange). (c) Eq. (\ref{eq:free-evolution}) with a random choice of the phases $\varphi_{n}$ (blue, dashed line). (d) Evolution of the quantity $R(t)$ in Eq. (\ref{eq:rx}), compared with the case of Eq. (\ref{eq:free-evolution}) for a random choice of the phases $\varphi_{n}$ (dotted line, $R\simeq0.959$).}
\label{fig:speckles}
\end{figure} 

Here we consider the quantum evolution of an initial Thomas-Fermi (TF) wave packet with the macroscopically occupied components lying below the effective mobility edge, as considered in \cite{sp,billy,succi,piraud}. We choose $V_{0}=0.19\,E_{\xi}$ and an initial TF radius $R_{TF}=340\, \xi$ \cite{TF}. 
The corresponding asymptotic density distribution and the evolution of the quantity $R(t)$ are shown in Fig. \ref{fig:speckles}. Again, in (a) we plot the asymptotic density as obtained from the full expression in Eq. (\ref{eq:free-evolution}), that has to be compared with the expression in Eq. (\ref{eq:asymptotic_density}) containing just the diagonal terms, shown in (b), and 
with the expression in Eq. (\ref{eq:free-evolution}) but with a random choice of the phases, in (c).
We find that the tails of the density distribution can be nicely fitted with the function $n_0(x)\propto|x|^{\alpha}e^{-2\gamma|x|}$, as discussed in \cite{sp,succi} (here we find $\alpha\approx2$, $\gamma\approx4\cdot 10^{-4}$). 
The same fitting lines are reported in all the three panels (a), (b) and (c), for comparison. These figures confirm that the phase randomization anstaz holds also for the case of the speckle potential, and that the asymptotic density distribution can be reproduced by the diagonal term in Eq. (\ref{eq:asymptotic_density}). Correspondingly, also the evolution of $R(t)$ in Fig. \ref{fig:speckles}(d) shows a behavior similar to the quasiperiodic case. Finally, we notice that a similar behavior holds also in case of a single speckle realization, but with larger local density fluctuations with respect to the average case.

\textit{Conclusions.}
We have shown that a convenient way to deal with the quantum localization of a wave packet in the presence of disorder is to consider its projection over the eigenstates of the disordered potential.
With this approach, the tails of the asymptotic density distribution are determined by the diagonal term of the density, that is independent of time \cite{capeta}. This is equivalent of assuming a randomization of the phase of the eigenstates, and corresponds to the fact that the off-diagonal (interference) terms contribute just with local fluctuations around the diagonal term. As specific examples, we have considered the cases of two recent experiments where Anderson localization was observed for ultracold atoms in speckle \cite{billy} and quasiperiodic potentials \cite{roati}. In the latter case we have also shown that the exponential decay of the tails in the continuum description is intrinsically limited by the presence of extended states in excited Bloch bands. The mechanism discussed in this paper provides a complementary approach to the usual description in terms of the localization of single momentum components as a result of multiple scattering from the random potential.

\begin{acknowledgments} 
We thank F. Dalfovo, G. Modugno, and L. Sanchez-Palencia for valuable comments and suggestions. We also acknowledge fruitful discussions with D. Clem\'ent, M. Larcher, and all the people in the QDG group at LENS in Florence.
\end{acknowledgments}


\begin{thebibliography}{}
\bibitem{numerics}
{We use the following prescription: ${H}_{ij} = (\hbar^{2}/2m\Delta x^2)(2\delta_{ij} -\delta_{i+1,j}-\delta_{i-1,j}) + V(j\Delta x) \delta_{ij}$, with vanishing boundary conditions, $\psi_0=\psi_{N+1}=0$. We take a system of length $L = 3200\xi$ and $N=16384 (=2^{14})$ grid points, with $\Delta x=L/N$.}
\bibitem{offdiag}
We note that $n_{off}$ is never vanishing, its amplitude being always of the order of the diagonal term. However, in the asymptotic regime it contributes just with local fluctuations around the diagonal term (in logarithmic scale).
\bibitem{DJ}
The critical time increases as $\Delta/J$ is decreased; in the diffusive regime, $\Delta/J<2$, the interference terms contribute substantially to the packet evolution and the phase randomization ansatz is no longer valid, as expected.
\bibitem{tenbands}
{For numerical convenience, here we have considered up to ten bands. We note that the same behaviour can be obtained even with just two bands.}
\bibitem{TF}
This value corresponds to $R_{TF}=7.96\, \xi$ for the trapped interacting BEC, before the pre-expansion stage where most of the interaction energy is transformed into kinetic energy, considered in \cite{sp,billy}.
\end{thebibliography}
\end{document}